\documentclass[prl,twocolumn,showpacs,superscriptaddress]{revtex4}

\usepackage{amsmath}
\usepackage{graphicx}

\usepackage{amsmath}
\usepackage{amssymb}
\usepackage{graphicx}
\usepackage[british]{babel}

\begin{document}

\preprint{\today.}

\title{Leapover lengths and first passage time statistics for L{\'e}vy flights}

\author{Tal Koren}
\affiliation{School of Chemistry, Tel Aviv University, Tel Aviv 69978, Israel}
\author{Michael A. Lomholt}
\affiliation{Physics Department, Technical University Munich, 85748 Garching,
Germany}
\author{Aleksei V. Chechkin}
\affiliation{Institute for Theoretical Physics NSC KIPT,
Akademicheskaya st.1, 61108 Kharkov, Ukraine}
\author{Joseph Klafter}
\affiliation{School of Chemistry, Tel Aviv University, Tel Aviv 69978, Israel}
\author{Ralf Metzler}
\affiliation{Physics Department, Technical University Munich, 85748 Garching,
Germany}

\begin{abstract}
Exact results for the first passage time and leapover statistics of symmetric
and one-sided L{\'e}vy flights (LFs) are derived.
LFs with stable index $\alpha$ are shown to have leapover lengths, that are
asymptotically power-law distributed with index $\alpha$ for one-sided LFs
and, surprisingly, with index $\alpha/2$ for symmetric LFs. The first passage
time distribution scales like a power-law with index $1/2$ as required by the
Sparre Andersen theorem for symmetric LFs, whereas one-sided LFs have a
narrow distribution of first passage times. The exact analytic results are confirmed by extensive simulations.
\end{abstract}


\maketitle

\begin{figure}
\includegraphics[width=8.6cm]{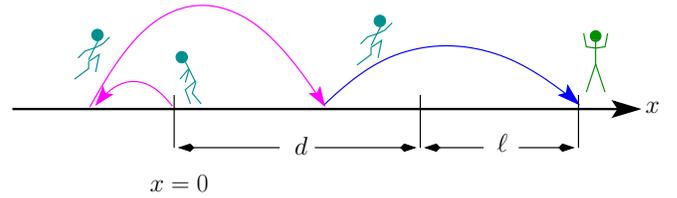}
\caption{Schematic of the leapover problem: the random walker starts at
$x=0$ and after a number of jumps crosses the point $x=d$, overshooting
it by a distance $\ell$.
}
\label{fig0}
\end{figure}

The statistics of first passage times is a classical concept to quantify
processes, in which it is of interest when the dynamic variable crosses
a given threshold value for the first time, e.g., when a tracer in some
aquifer reaches a certain probe position, two molecules meet to form a
chemical bond, animals search for sparse food locations, or a share at the
stock market crosses a preset market value \cite{redner,hughes,vankampen}.
Here, we revisit the first passage time problem for processes with non-trivial
jump length distributions, namely, L{\'e}vy flights (LFs) and derive exact
asymptotic expressions for the first passage time density $p_f(\tau)$ of
symmetric and one-sided LFs. For the former, we obtain the Sparre Andersen
universality $p_f(\tau)\simeq \tau^{-3/2}$, while a narrow behavior is
found for one-sided LFs. Apart from calculating the first passage times, we
investigate the behavior of the first passage leapovers, that is, the distance
the random walker overshoots the threshold value $d$ in a single jump
(see Fig.~\ref{fig0}). Surprisingly, for symmetric LFs with jump length
distribution $\lambda(x)\simeq |x|^{-1-\alpha}$ ($0<\alpha<2$) the distribution
of leapover lengths across $x=d$ is distributed like $p_l(\ell)\simeq \ell^
{-1-\alpha/2}$, i.e., it is much broader than the original jump length
distribution. In contrast, for one-sided LFs the scaling of $p_l
(\ell)$ bears the same index $\alpha$.

For processes subject to a narrow jump length distribution with finite second
moment $\int_{-\infty}^{\infty}x^2\lambda(x)dx$ the crossing of a given
threshold value $d$ is identical to the first arrival at $x=d$ \cite{hughes}.
This is no longer true for LFs: Intuitively, a particle, whose jump lengths
are distributed according to the symmetric long-tailed distribution $\lambda
(x)\simeq|x|^{-1-\alpha}$  ($0<\alpha<2$) is likely to criss-cross the point
$x=d$ multiple times before it eventually hits it, causing the first arrival
at $d$ to be slower than its first passage across $d$ \cite{fpt}. A
measure for the ability to criss-cross $d$ is the distribution of leapover
lengths, $p_l(\ell)$. Information on the leapover behavior is therefore
important to the understanding of how far proteins searching for their
specific binding site along DNA overshoot their target \cite{michael},
climatic forcing visible in ice core records exceeds a given value
\cite{ditlevsen}, or to define better stock market
strategies determining when to buy or sell a certain share instead of fixing
a threshold price \cite{stock}. The quantification of leapovers is
vital to estimate how far diseases would spread once a carrier of that
disease crosses a certain border \cite{brockmann}. Leapover statistics of
one-sided LFs provide an interesting alternative interpretation of the
distribution of the first waiting time in ageing continuous time random
walks \cite{eli}, just to name a few examples.

The master equation for a Markovian diffusion process,
\begin{eqnarray}
\nonumber
\frac{\partial P(x,t)}{\partial t}=\frac{1}{\tau}\int_{-\infty}^{\infty}
&&\left[\lambda(x-x')P(x',t)\right.\\
&&\left.-\lambda(x'-x)P(x,t)\right]dx'
\label{me}
\end{eqnarray}
accounts for the influx of probability to position $x$, and the outflux away
from $x$, where $\lambda(x)$ is a general, normalized jump length distribution.
The time scale for single jumps is $\tau$. The solution to Eq.~(\ref{me}) in
Fourier space is $P(k,t)=e^{-\left[1-\lambda(k)\right]t/\tau}$,
denoting the Fourier transform $f(k)=\int_{-\infty}^{\infty}e^{ikx}f(x)
dx$ by explicit dependence on the wave number $k$. For instance, for the
symmetric jump length distribution $\lambda(x)\simeq\sigma^{\alpha}|x|^{-1
-\alpha}$, one finds
\begin{equation}
\label{symm}
P(k,t)=e^{-K^{(\alpha)}|k|^{\alpha}t}
\end{equation}
with $K^{(\alpha)}=\sigma^{\alpha}/\tau$, the characteristic function of a
symmetric L{\'e}vy stable law as obtained from continuous time random walk
theory in the diffusion limit or from the equivalent space fractional
diffusion equation \cite{report}.

In the following we study processes with the long-tailed composite jump
length distribution
\begin{equation}
\label{jld}
\lambda(x)/\tau=\Theta(|x|-\varepsilon)\left[c_1\Theta(-x)+c_2\Theta(x)\right]/
|x|^{1+\alpha},
\end{equation}
where $\Theta(x)$ is the Heaviside function. For $c_1=c_2$, $\lambda(x)$
defines a symmetric LF, and for $c_1=0$ and $c_2>0$ a completely
asymmetric (one-sided) LF permitting exclusively forward jumps. The cutoff
$\varepsilon$ excludes the singularity at $x=0$,
but can be taken to be small, $\varepsilon\to 0$ \cite{REM}.

In the theory of homogeneous random processes with independent jumps there
exists a theorem, which provides an exact expression for the joint PDF $p(\tau,
\ell)$ of first passage time $\tau$ and leapover length $\ell$ ($\ell\ge0$)
across $x=d$ for a particle initially seeded at $x=0$ \cite{gikhman,skorokhod}.
We here evaluate this theorem, that appears to have been widely overlooked,
and derive a number of new analytic results for $p_f(\tau)$ and $p_l(\ell)$ of
symmetric and one-sided LFs. With the probability to jump longer than $x$,
\begin{equation}
\mathcal{M}(x)=\int_x^{\infty}\lambda(x')dx',\quad x>0,
\end{equation}
the theorem states that the double Laplace transform of the joint PDF
\cite{gikhman,skorokhod}
\begin{equation}
p(u,\mu)=\int_0^{\infty}\int_0^{\infty}e^{-u\tau-\mu\ell}p(\tau,
\ell)d\tau d\ell
\end{equation}
is given in terms of the multiple integral
\begin{eqnarray}
\nonumber
p(u,\mu)&=&1-q_+(u,d)-\frac{\mu}{u}\int_0^d\frac{\partial q_+(u,
s)}{\partial s}ds\\
\nonumber
&&\times\int_{-\infty}^0\frac{\partial q_-(u,s')}{\partial s'}d
s'\int_0^{\infty}e^{-\mu s''}\\
&&\times \mathcal{M}(d+s''-s'-s)ds''.
\label{multint}
\end{eqnarray}
Here, we use the two auxiliary measures
$q_{\pm}(u,x)$ defined through Fourier transforms
\begin{multline}
{\tilde q}_\pm(u,k)=\int_{-\infty}^\infty e^{i k x}\frac{\partial q_\pm(u,x)}{\partial x}d x \\
=\exp\left\{\pm\int_0^{\infty}\frac{e^{-ut}}{t}\int_0^{\pm\infty}
\left(e^{ikx}-1\right)P(x,t)dxdt\right\},
\label{measure}
\end{multline}
and the condition $q_\pm(u,0)=0$. They are related to the cumulative
distributions of the maximum, $Q_+(t,d)=\mathrm{Pr}\left\{\max_{0\le\tau\le
t} x(\tau)<d \right\}$, and minimum, $Q_-(t,d)=\mathrm{Pr}\left\{\min_{0\le
\tau\le t} x(\tau)<d\right\}$, of the position $x(t)$ such that $q_\pm (u,d)
=u\int_0^\infty e^{-u t}Q_\pm(t,d)d t$.
The complicated integrals above reduce to elegant results for
symmetric and one-sided LFs, as we show now.

\begin{figure}
\includegraphics[width=8.6cm]{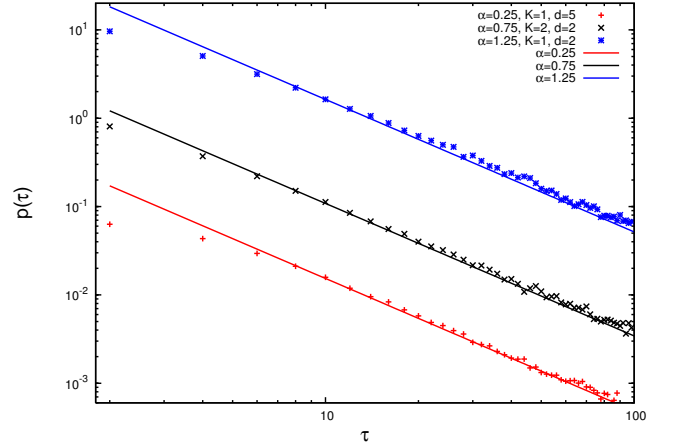}
\caption{First passage time density $p_f(\tau)$ for symmetric LFs with Sparre
Andersen universality $p_f(\tau)\simeq \tau^{-3/2}$. The curves for $\alpha=
0.75$ and 1.25 are multiplied by a factor 10 and 100, for better
visibility. Lines: theory. Symbols: simulations.}
\label{fig1}
\end{figure}

For \emph{symmetric LFs\/} ($c_1=c_2\equiv c$), the propagator is defined by
the characteristic function (\ref{symm})
with generalized diffusion coefficient
$K^{(\alpha)}=2c\Gamma(1-\alpha)\cos(\pi\alpha/2)/\alpha$.
In the limit $u\to 0$ (long time limit), we obtain from Eq.~(\ref{measure})
\begin{equation}
{\tilde q}_+(u,k)\sim\frac{u^{1/2}}{\sqrt{K^{(\alpha)}}|k|^{\alpha/2}}
\exp\left\{\frac{i\mathrm{sign}(k)\pi\alpha}{4}\right\}.
\end{equation}
Inverse Fourier transform yields
\begin{equation}
q_+(u,d)\sim\frac{2u^{1/2}}{\alpha\sqrt{K^{(\alpha)}}\Gamma(\alpha/2)}
d^{\alpha/2},\quad d>0
\end{equation}
such that from $p_f(u)=1-q_+(u,d)$ we find
\begin{equation}
p_f(\tau)\sim\frac{d^{\alpha/2}}{\alpha\sqrt{\pi K^{(\alpha)}}\Gamma(
\alpha/2)}\tau^{-3/2}.
\end{equation}
This is the exact asymptotic first passage time PDF of
symmetric LFs. Fig.~\ref{fig1} demonstrates good agreement with simulations
results, for which the algorithm from Ref.~\cite{chambers} was used to obtain
random numbers distributed according to L{\'e}vy stable laws. We note that
previously only the $\tau^{-3/2}$ scaling was known from simulations and
application of the Sparre Andersen theorem \cite{fpt}.

\begin{figure}
\includegraphics[width=8.8cm]{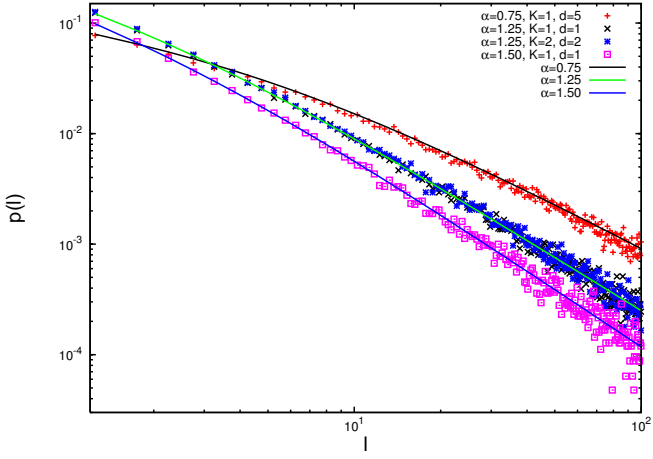}
\caption{Leapover density $p_l(\ell)$ for symmetric LFs.}
\label{fig2}
\end{figure}

For symmetric LFs, for $0<\alpha<2$ we obtain that
\begin{equation}
\label{em}
\mathcal{M}(x)=\frac{K^{(\alpha)}}{2\Gamma(1-\alpha)\cos(\pi\alpha/2)}x^{-\alpha},\,\,
\,\,x>0.
\end{equation}
Using that for symmetric LFs $q_-(\tau,x)=q_+(\tau,-x)$ it turns out
after some transformations from Eq.~(\ref{multint}) that
\begin{equation}
p_l(\mu)=\int_0^{\infty}e^{-\mu \ell}\frac{\sin(\pi\alpha/2)}{\pi}\frac{
(d/\ell)^{\alpha/2}}{d+\ell}d \ell,
\end{equation}
from which it follows immediately that
\begin{equation}
p_l(\ell)=\frac{\sin(\pi\alpha/2)}{\pi}\frac{d^{\alpha/2}}{\ell^{\alpha/2}(d+
\ell)},
\end{equation}
see Fig.~\ref{fig2}.
Note that $p_l$ is normalized. In the limit $\alpha\to2$, $p_l(\ell)$ tends
to zero if $\ell\neq0$ and to infinity at $\ell=0$ corresponding to the
absence of leapovers in the Gaussian continuum limit. However, for $0<\alpha
<2$ the leapover PDF follows an asymptotic power-law with index $\alpha/2$,
and is thus broader than the original jump length PDF $\lambda(x)$ with index
$\alpha$. This is a remarkable finding: while $\lambda$ for $1<\alpha<2$ has
a finite characteristic length $\langle|x|\rangle$, the mean leapover length
diverges.

\begin{figure}
\includegraphics[width=8.6cm]{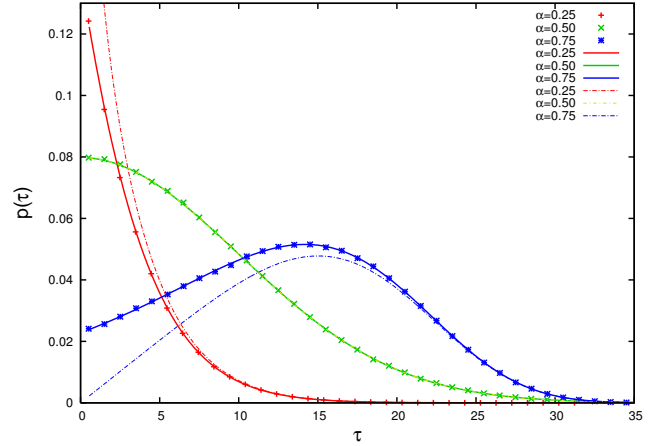}
\caption{First passage density for one-sided LF ($K^{(\alpha)}=1$). The thick
lines represent numerical evaluations of the exact analytic expression, while
the thin dashed lines correspond to the asymptotic behavior (\ref{masymp}).
Symbols: simulations.}
\label{fig3}
\end{figure}

Consider now \emph{one-sided\/} LFs with $c_1=0$ in
Eq.~(\ref{jld}). In this case, the PDF has the characteristic function
\begin{equation}
P(k,t)=\exp\left\{-K^{(\alpha)}t|k|^{\alpha}\left[1-i\mathrm{sign}(k)\tan
\left(\frac{\pi\alpha}{2}\right)\right]\right\},
\end{equation}
where $K^{(\alpha)}=c_2\Gamma(1-\alpha)\cos(\pi\alpha/2)/\alpha$ and
$\mathcal{M}(x)$ for $x>0$ is twice the expression in Eq.~(\ref{em}).
Eq.~(\ref{measure}) leads to
\begin{equation}
\label{qnew}
{\tilde q}_+(u,k)=\frac{u}{u+\zeta},\,\,\zeta=K^{(\alpha)}(-ik)^{\alpha}/
\cos\left(\frac{\pi\alpha}{2}\right),
\end{equation}
as $(-ik)^{\alpha}=[-i\mathrm{sign}(k)|k|]^{\alpha}=|k|^{\alpha}\exp[-
i\mathrm{sign}(k)\pi\alpha/2]$. From this we calculate that the Fourier
transform of $\langle\exp(-u\tau)\rangle=\int_0^{\infty}\exp(-u\tau)p_f(\tau)
d\tau$ can be written as
\begin{equation}
\int_{-\infty}^{\infty}e^{ikx}\langle e^{-u\tau}\rangle dx
=\frac{(-ik)^{\alpha-1}}{(-ik)^{\alpha}+u\cos(\pi\alpha/2)/K^{(\alpha)}},
\end{equation}
and we change the variable $ik\to-s$ to find \cite{leapoverlong}
\begin{equation}
\label{mittag1}
\langle e^{-u\tau}\rangle=E_{\alpha}\left[-\frac{u}{K^{(\alpha)}}\cos
\left(\frac{\pi\alpha}{2}\right)d^{\alpha}\right].
\end{equation}
Here, we used the definition of the Mittag-Leffler function
\begin{equation}
\int_0^{\infty}E_{\alpha}\left(-\theta x^{\alpha}\right)e^{-sx}dx=
\frac{s^{\alpha-1}}{s^{\alpha}+\theta}.
\end{equation}
whose series expansion and asymptotic behavior are \cite{report}
\begin{equation}
\label{mittag2}
E_{\alpha}(-z)=\sum_{n=0}^{\infty}\frac{(-z)^n}{\Gamma(1+\alpha n)}
\sim\sum_{n=0}^{\infty}\frac{(-1)^nz^{-1-n}}{\Gamma(1-\alpha
[n+1])}.
\end{equation}
From the relation between $E_{\alpha}$ and the $M_{\alpha}$-function
\cite{mainardi},
\begin{equation}
\int_0^{\infty}e^{-ut}M_{\alpha}(t)dt=E_{\alpha}(-u),\,\,0<\alpha<1,
\end{equation}
the following result for the first passage time PDF yields
\begin{equation}
p_f(\tau)=\frac{K^{(\alpha)}}{\cos\left(\alpha\pi/2\right)d^\alpha}M_\alpha\left(
\frac{K^{(\alpha)}\tau}{\cos\left(\alpha\pi/2\right)d^\alpha}\right)\;.
\end{equation}
The $M_{\alpha}$-function has the series representation
and asymptotic behavior with exponential decay
\begin{eqnarray}
M_{\alpha}(z)=\sum_{n=0}^{\infty}\frac{(-z)^n}{n!\Gamma(1-\alpha-\alpha n)}
\hspace*{3.0cm}\\
\label{masymp}
\sim \frac{(\alpha z)^{(\alpha-1/2)/(1-\alpha)}}{\sqrt{2\pi(1-
\alpha)}}\exp\left[-\frac{1-\alpha}{\alpha}(\alpha z)^{1/(1-\alpha)}\right]\;.
\end{eqnarray}
The moments of the $M_{\alpha}$-function are obtained through
\begin{equation}
\int_0^{\infty}z^nM_{\alpha}(z)dz=\lim_{s\to0}(-1)^n\frac{d^n}{ds^n}E_{\alpha}
(-s)=\frac{\Gamma(n+1)}{\Gamma(1+\alpha n)},
\end{equation}
from which we calculate the mean first passage time
\begin{equation}
\langle\tau\rangle=\frac{d^\alpha\cos(\pi\alpha/2)}{K^{(\alpha)}\Gamma(1+
\alpha)},
\end{equation}
that is \emph{finite\/} and
grows with the $\alpha$th power of the distance $d$. For $\alpha=1/2$, we
recover the exact form \cite{iddoyossi}
\begin{equation}
p_f(\tau)=K^{(\alpha)}\sqrt{\frac{2}{\pi d}}\exp\left(-\frac{\left(K^{(\alpha)}
\right)^2\tau^2}{2d}\right).
\end{equation}
The first passage PDF $p_f(\tau)$ is
displayed in Fig.~\ref{fig3} in nice agreement with the simulations. Note that
for $\alpha\le1/2$ the tail of $\lambda(x)$ is so long that it is most
likely to cross $x=d$ in the first jump, while for $\alpha>1/2$, $p_f(\tau)$
has a maximim at finite $\tau>0$.

To obtain the leapover statistics for the one-sided LF, we first note that
since $P(x<0,t)=0$ (only forward steps are permitted) we have $q_-(u,k)=1$,
and thus $\partial q_-(u,x)/\partial x=\delta(x)$. Combining Eqs.~(\ref{multint}) and
(\ref{measure}),
\begin{eqnarray}
\nonumber
\langle e^{-\mu\ell}\rangle=1-\lim_{u\to 0}\frac{\mu}{u}\int_0^d\int_0^{\infty}
e^{-\mu s'}\mathcal{M}(d+s'-s)\\
\times\frac{\partial q_+(u,s)}{\partial s}d s'd s.
\label{rel}
\end{eqnarray}
With the small $u$ expansion of the Mittag-Leffler function,
Eqs.~(\ref{mittag1}) and (\ref{mittag2}) produce
\begin{equation}
\label{mittag3}
\frac{\partial q_+(u,x)}{ \partial x}=\frac{u\cos(\pi\alpha/2)}{K^{(\alpha)}\Gamma(\alpha)}
x^{\alpha-1}.
\end{equation}
Eqs.~(\ref{qnew}) and (\ref{mittag3}) inserted into Eq.~(\ref{rel}) then yield
\begin{equation}
p_l(\mu)=\langle e^{-\mu\ell}\rangle=\frac{\sin(\pi\alpha)}{\pi}\int_0^{\infty}
e^{-\mu\ell}\frac{d^{\alpha}}{\ell^{\alpha}(d+\ell)},
\end{equation}
leading to the leapover PDF
\begin{equation}
\label{lod}
p_l(\ell)=\frac{\sin(\pi\alpha)}{\pi}\frac{d^{\alpha}}{\ell^{\alpha}(d+\ell)},
\end{equation}
which corresponds to the result obtained in Ref.~\cite{iddoyossi} from a
different method. Thus, for the one-sided LF, the scaling of the leapover
is exactly the same as for the jump length distribution, namely, with exponent
$\alpha$.

\begin{figure}
\includegraphics[width=8.6cm]{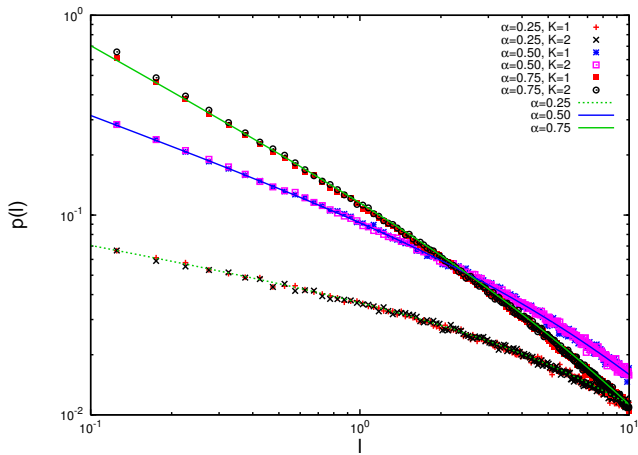}
\caption{Leapover distribution for one-sided LF with $d=10$.}
\label{fig4}
\end{figure}

The leapover distribution (\ref{lod}) also provides a new aspect to the
first waiting time in a renewal process with broad waiting time distribution
$\psi(t)\simeq t^{-1-\beta}$ ($0<\beta<1$). 
Interprete the position $x$ as time and the
jump lengths drawn from the one-sided $\lambda(x)$ as waiting times $t$.
Consider an experiment, starting at time $t_0$, on a system prepared at
time 0 (corresponding to position $x=0$). Then the first recorded waiting
time $t_1$ of the system will be distributed like $p_1(t_1)=\pi^{-1}\sin(\pi
\alpha)t_0^{\alpha}/[t_1^{\alpha}(t_0+t_1)]$, as obtained from a different
reasoning in Ref.~\cite{eli}. We note that the first passage time $\tau$
in this analogy corresponds to the number of waiting events.


While for symmetric LFs it was previously established that the first passage
time distribution follows the universal Sparre Andersen asymptotics $p_f(\tau)
\simeq\tau^{-3/2}$, here we derived the prefactor of this law, in particular,
its dependence on the generalized diffusion coefficient $K^{(\alpha)}$. For
the same case, we derived the leapover distribution $p_l(\ell)$, that is
surprising for two reasons: first, $p_l(\ell)$ is independent of $K^{(\alpha)}$,
synonymous to the noise strength; and second, its power-law exponent is
$\alpha/2$, and thus $p_l(\ell)$ is broader than the original jump length
distribution.

For one-sided LFs, we recovered the previously reported leapover distribution
and derived the so far unknown first passage time distribution. While the
leapovers follow the same asymptotic scaling $p_l(\ell)\simeq\ell^{-1-\alpha}$
as the jump lengths $\lambda(x)$, once more independent of $K^{(\alpha)}$, the
first passage times are narrowly distributed. We also drew
an analogy between the leapovers and the first waiting time in a subdiffusive
renewal process. For both symmetric and one-sided LFs, extensive simulations
showed nice agreement with the theoretical results, without adjustable
parameters.

We acknowledge partial funding from NSERC and the Canada Research Chairs
programme.


\begin{thebibliography}{99}
\bibitem{REM} The same trick is used in computer simulations of LFs. In a
mathematical sense, this means that $\tau\to0$ such that $u\tau\ll1$ is
fulfilled.
\end{thebibliography}
\end{document}